\documentclass[twocolumn,reprint,nofootinbib,longbibliography]{revtex4-1}
\usepackage{amsfonts, amsmath, amssymb, bm, enumerate, graphicx, graphics, color,mathrsfs,hyperref}

\newcommand{\Lie}[0]{{\cal L}\, }

\newcommand{\tl}{\theta_{(\ell)}}
\newcommand{\tn}{\theta_{(n)}}

\newcommand{\nn}{\nonumber}
\newcommand{\be}{\begin{equation}}
\newcommand{\ee}{\end{equation}}
\newcommand{\bea}{\begin{eqnarray}}
\newcommand{\eea}{\end{eqnarray}}

\newcommand{\cV}{\mathcal{V}}

\newcommand{\subs}[1]{_{\! \! \mbox{\tiny{ #1}}}}     
\newcommand{\sups}[1]{^{\! \! \mbox{\tiny{ #1}}}}     

\begin{document}

\title{%Vaidya Reissner-Nordstr\" om spacetimes: 
Evolutions from extremality}

\date{\today}
\author {Ivan Booth}
\email{ibooth@mun.ca}
\affiliation{
Department of Mathematics and Statistics\\ Memorial University of Newfoundland \\  
St. John's, Newfoundland and Labrador, A1C 5S7, Canada \\
}

\begin{abstract}
We examine the evolution of extremal spherically symmetric black holes, developing both general theory 
as well as the specific cases of (charged) null dust and massless scalar field spacetimes. 
As matter accretes onto extremal marginally trapped tubes, they generically evolve to become non-extremal with 
the initial extremal horizon bifurcating into inner and outer non-extremal horizons. 
At the start of this process arbitrarily slow matter accretion can cause a geometrically invariant measure of horizon growth 
to jump from zero to infinity. We also consider dynamical horizons that are extremal throughout their evolution and see that such spacetimes
contain two extremal black hole horizons: an inner isolated one and an outer dynamical one. We compare these extremal dynamical horizons
 with the dynamical extreme event horizon spacetimes
of Murata, Reall and Tanahashi.
\end{abstract}

\maketitle

\section{Introduction}
Extremality plays an important role in the mathematics and physics of black holes. Thermodynamically, extremal black holes are
zero temperature states and subject to the third law of black hole mechanics: physical processes cannot turn a non-extremal
hole into an extremal one\cite{Israel_ThirdLaw}. Supersymmetric black holes are necessarily extremal and this made the first
string theory calculations of black hole entropy possible\cite{Strominger:1996sh}. Physically, for any given horizon area, black holes have 
a maximum charge and rotation and those bounds are saturated by extremal holes\cite{Hennig:2008yw,Mars:2012sb,Jaramillo:2011pg,Clement:2012vb}. 
Mathematically their horizon geometry is tightly constrained in any dimension and indeed in four-dimensions 
the intrinsic geometry of any extremal horizon in Einstein-Maxwell theory is isometric to a member of the Kerr-Newman family\cite{Lewandowski:2002ua}.
Through the near horizon formalism, the tight constraints have also enabled great progress in the classification of 
(time-independent) extremal black hole horizons in five and higher dimensions \cite{Kunduri:2013gce}.

Most of the work mentioned in the preceding paragraph has focussed on stationary extremal black holes. However it has recently 
been shown that these very important solutions are unstable and so not expected to remain stationary under generic conditions. 
The initial proof of the instability of extremal Reissner-Nordstr\"om black holes
\cite{Aretakis:2011ha,Aretakis:2011hc} was quickly extended to Kerr-Newman black holes \cite{Aretakis:2012bm,Aretakis:2012ei,Lucietti:2012sf}. Subsequent work has included  numerical probes of that instability \cite{Lucietti:2012xr,Murata:2013daa}. 

Given these instabilities it is of interest to understand the evolution of initially extremal black holes and in this paper we study 
both exits from extremality and special cases of dynamical  extremal horizons. Somewhat surprisingly such evolutions have
not received much attention in the literature. As such there are simple yet still interesting cases that have not been studied
and we tackle one of them here.  We restrict our attention to spherically symmetric spacetimes and focus our attention mainly on the evolution 
of the marginally trapped tubes (essentially apparent horizons) though we will also briefly consider event horizons. 
We primarily use Vaidya Reissner-Nordstr\"om (VRN) spacetimes as concrete examples but also briefly consider massless scalar fields.

This paper is organized as follows.  Section \ref{background} develops the general theory of marginally trapped tubes in 
spherically symmetric backgrounds. It sets up notation and reviews two-surface geometry both in general and as applied to
geometric horizons. It then considers the kinematics and dynamics of those horizons. 
Section \ref{nulldust} considers the special case of null dust accreting onto an existing black hole. It reviews Vaidya Reissner-Nordstr\"om
and then examines both transitions from extremality and evolutions at extremality. Section \ref{scalar} explores some of 
the results of Murata, Reall and Tanahashi \cite{Murata:2013daa}, first considering the interaction of extremal horizons and massless scalar fields
and then examining  dynamical extremal event horizons. Finally Section \ref{discuss} reviews and discusses our results.

\section{Background and general theory}
\label{background}

In this section we establish the background material that we will need for the calculations: the geometry of two-surfaces, definitions of geometric horizons and  the kinematics 
and dynamics of marginally trapped tubes. 
Throughout we restrict our attention to spherically symmetric spacetimes and marginally trapped tubes. 
For more details on the geometry, using the same notation (including more general non-symmetric cases) see \cite{Booth07}.

\subsection{Spacetime and two-surface geometry}
\label{geometry}

Let $(M, g_{ab}, \nabla_a)$ be a $(3+1)$-dimensional spherically symmetric spacetime. By the symmetry it may be 
decomposed into spacelike two-surfaces $S{(t,r)}$ where $t$ is a time parameter and  $r$ is the areal radius of the surface. 

There are four null directions from each surface and we assume that we are working in a spacetime in which those directions may be identified as future/past and inwards/outwards.
The spherically symmetric vectors $\ell$ and $n$ respectively point in the 
future outward and future inward directions and are cross-normalized so that $\ell \cdot n = -1$. 
This leaves one degree of rescaling freedom $\ell \rightarrow e^{f} \ell$ and $n \rightarrow e^{-f} n$
in the null vectors. We assume that the scaling is also spherically symmetric so that the allowed freedom is given by $f=f(t,r)$. 

The induced metric on the two-surfaces is
\be
\tilde{q}_{AB} dx^A dx^B = r^2 d \Omega^2 
\ee
where $d\Omega^2$ is the usual metric on the unit two-sphere while their outward and inward expansions are respectively:
\bea
\theta_{(\ell)} & = & %\tq^{ab} \nabla_a \ell_b 
 \frac{1}{\sqrt{\tilde{q}}} \Lie_\ell \sqrt{\tilde{q}} = 2 \Lie_\ell \ln (r)  \; \; \mbox{and} \label{tl} \\ 
\theta_{(n)} & = & % \tq^{ab} \nabla_a n_b 
 \frac{1}{\sqrt{\tilde{q}}} \Lie_n \sqrt{\tilde{q}} = 2 \Lie_n \ln (r) \, ,
\eea
where $\Lie_\ell$ indicates a Lie derivative and $\sqrt{\tilde{q}}=r^2 \sin  \theta$ is the area element on the $S(t,r)$. These expansions
are used to classify surfaces. In particular we will be interested in \emph{untrapped} surfaces ($\tl > 0$, $\tn < 0$), \emph{trapped} surfaces ($\tl< 0, \tn <0$), \emph{marginally outer trapped}
surfaces (MOTS) ($\tl = 0$) and \emph{marginally trapped} surfaces (MTS) ($\tl = 0, \tn< 0$).

In later sections we will need to know the rates of change of these expansions. These are:
\bea
\Lie_\ell \tl -  \kappa_{\ell} \tl&=& -G_{ab} \ell^a \ell^b - \frac{1}{2} \tl^2 \label{Lltl} \\
\Lie_n \tl + \kappa_{n} \tl &=&- \tilde{K} + G_{ab} \ell^a n^b - \tl \tn  \label{Lntl} \\
\Lie_\ell \tn + \kappa_{\ell} \tn &=& - \tilde{K} + G_{ab} \ell^a n^b -  \tl \tn  \label{Lltn}\\
\Lie_n \tn - \kappa_{n} \tn &=&-G_{ab} n^a n^b - \frac{1}{2} \tn^2  \label{Lntn}
\eea
where $\tilde{K} = 1/r^2$ is the Gauss curvature of $S(t,r)$ and $\kappa_X = - X^a n_b \nabla_a \ell^b$. 

\subsection{MOTTs: definition and  classification}
\label{horizons}

There are several closely related geometrical notions of horizon including apparent horizons \cite{hawking73}, trapping horizons 
\cite{Hayward:1993wb}, isolated/dynamical horizons \cite{Ashtekar:2004cn}, marginally trapped tubes \cite{Ashtekar:2005ez}  and, most 
recently, future holographic screens \cite{Bousso:2015mqa}. A summary of the definitions of these
 various objects and their complex nomenclature may be found in \cite{Booth:2005qc}\footnote{Except for holographic screens
which, though viewed from a different physical perspective, are mathematically identical to marginally trapped tubes.}.

In all of these notions horizons are \emph{marginally outer trapped tubes} (MOTTs): three-surfaces $H$ foliated by MOTS. \emph{Isolated
horizons} are null and time-independent MOTTs (with tangent $\ell$). Essentially they are equilibrium states.  In this paper we 
usually further specialize to   \emph{marginally trapped tubes} (MTTs): MOTTs with $\tn < 0$. 

A typical black hole also has trapped surfaces inside the horizon and untrapped surfaces outside. That 
there be outer trapped surfaces ``just inside" a spherical MOTS\footnote{For general MOTS things are more complicated as it is necessary to allow for
rescaling of the null vectors. We can ignore those complications.} can be formalized as requiring
$
\Lie_n \tl < 0 \, .
$
In Hayward's classification \cite{Hayward:1993wb} this is the defining condition for an \emph{outer} trapping horizon. Researchers working
with initial data have a slightly stronger condition which they call  \emph{strictly stably outermost}  (introduced in 
\cite{Andersson:2005gq, Andersson:2007fh} and much used since). Given a MOTS with outward-pointing spacelike normal $\hat{r}$ 
in a Cauchy surface $\Sigma$ this condition requires
$
\Lie_{\hat{r}} \tl > 0 \, .
$
Such a vector can always be written as $\hat r = A \ell - B n$ for some positive $A$ and $B$. Then in spherical symmetry
\be
\Lie_{\hat{r}} \tl > 0 \Longrightarrow \Lie_n \tl < \left( \frac{B}{A} \right) \Lie_\ell \tl \, . 
\ee
Given the null energy condition $\Lie_\ell \tl < 0$ by (\ref{Lltl}) and so this implies that a strictly stably outermost MOTS is an outer trapping horizon. 
For isolated horizons the notions are equivalent. 

% In the literature
%this condition is given a couple of different names. Researchers working with initial data call marginally trapped surfaces
%which satisfy this condition \emph{strictly stably outermost} or some closely related phrase (see, for example, 
%\cite{Andersson:2005gq,Dain:2014bpa}). Equivalently in Hayward's classification of trapping horizons 
%\cite{Hayward:1993wb} these are \emph{outer} trapping horizons with $\Lie_n \tl < 0$. 

However, these conditions certainly do not hold for all MOTS. They are characteristic of outer 
horizons (hence Hayward's nomenclature)  but are violated for inner horizons. For example the
inner Cauchy horizon of a non-extremal Reissner-Nordstr\"om 
black hole has $\Lie_n \tl > 0$ with trapped surfaces outside and untrapped surfaces inside. 
%In Hayward's classification these are inner trapping horizons. 

For isolated MTTs the rate of change of the inward expansion is tied to the surface gravity. To summarize an extended discussion from \cite{Booth:2007wu} first note that
with a sensible choice of the scaling parameter so that $\Lie_\ell \tn =0$, by (\ref{Lntl}) and (\ref{Lltn}) we have
\be
\kappa_\ell \tn = \Lie_n \tl \, . \label{kappaLntl}
\ee
With $\tn <  0$, the surface gravity $\kappa_\ell$ vanishes if and only if $\Lie_n \tl = 0$. Note that $\Lie_n \tl$ for a spherical MOTS does not depend 
on the scaling of the null vectors. To see this set $\tl=0$ in (\ref{Lntl}).

Such time-independent MTTs are \emph{extremal} and 
we can rewrite this condition in a even more familiar form. Applying the  electromagnetic stress energy tensor
\be
T_{ab}^{\mbox{\tiny{EM}}} = \frac{1}{4\pi} \left(F_a^{\phantom{a} c} F_{bc} - \frac{1}{4} F_{cd} F^{cd} g_{ab} \right) \, , 
\label{TEM}
\ee  
we find that on a spherical MOTS:
\be
\Lie_n \tl = - \frac{1}{r^4} \left(r^2 - q^2 \right) + 8 \pi T^{\mbox{\tiny{nEM}}}_{ab} \ell^a n^b \, ,  \label{LntLS}
\ee
where $r$ is the areal radius of the surface, $q$ is the electric charge
contained by the surface (we assume the magnetic charge is zero), and $T^{\mbox{\tiny{nEM}}}_{ab}$ is the stress-energy tensor for any non-Maxwell fields. 
This last term vanishes for some matter fields (including the massless scalar fields and null dust that we will consider
later). In these cases
\be
\Lie_n \tl  \leq 0  \; \Longleftrightarrow \;  \kappa_\ell \geq 0 \;  \Longleftrightarrow \; r \geq q \, \,,  \label{DegExt}
\ee
with the saturation of one bound implying saturation of all the others.  For isolated horizons all of the common notions of extremality are equivalent. A spherically symmetric extremal isolated horizon has vanishing surface gravity,
no trapped surfaces directly inside and (for the matter models we consider) $r=q$. 

For time-dependent MTTs, $\Lie_\ell \tn$ should not vanish in general and so the vanishing $\kappa_\ell$ condition decouples from the other two. Physically this is not surprising
as it is not clear how surface gravity should be defined (or even if it can be) for non-stationary spacetimes. That said, the $\Lie_n \tl = 0 \; \Longleftrightarrow\;  r=q$ equivalence 
remains and it is this dual condition that we will use as a definition of (spherically symmetric) isolated and dynamical extremal  horizons. Again see \cite{Booth:2007wu} for 
a more detailed discussion. 

Finally note that while we assumed spherical symmetry in showing $r \geq q$ this is actually a much more general bound which holds for all strictly stably outermost  MOTS in 
spacetimes that satisfy the dominant energy condition (including time-dependent cases)\cite{Dain:2011kb}. 

\subsection{Dynamical MTTs: kinematics}
\label{kinematics}
In the classification we focussed on time-independent MTTs. We now consider dynamical ones in more detail. 
First, as long as the radial tangent vector to $H$ is not parallel to $n$ it may be written as
\be
\mathcal{V} =  \ell - C n \label{cV}
\ee
for an expansion parameter $C$ and appropriate scaling of the null vectors \cite{Booth:2003ji,Booth07}. 
The sign of $C$ determines the signature of $H$ and  with $\tl =0$,
\be
\theta_{(\cV)} = - C \tn \, .  
\ee
Thus for an MTT with:
\begin{description}
\item[$C>0$] $H$ is spacelike and expanding
\item[$C=0$] $H$ is null (tangent to $\ell$) and non-expanding 
\item[$C<0$] $H$ is timelike and contracting.
\end{description}
Using the language of \cite{Ashtekar:2004cn,Ashtekar:2005ez} these are respectively a \emph{dynamical horizon}, a \emph{(weakly) isolated horizon} and a 
\emph{timelike membrane}. As indicated by the terminology, an isolated horizon represents an equilibrium state. For example all Killing horizons
are isolated. 
If $H$ is tangent to $n$ then by (\ref{cV}), $C = \pm \infty$. We will refer to this as a \emph{null membrane}. 

\begin{figure}
\includegraphics[scale=0.65]{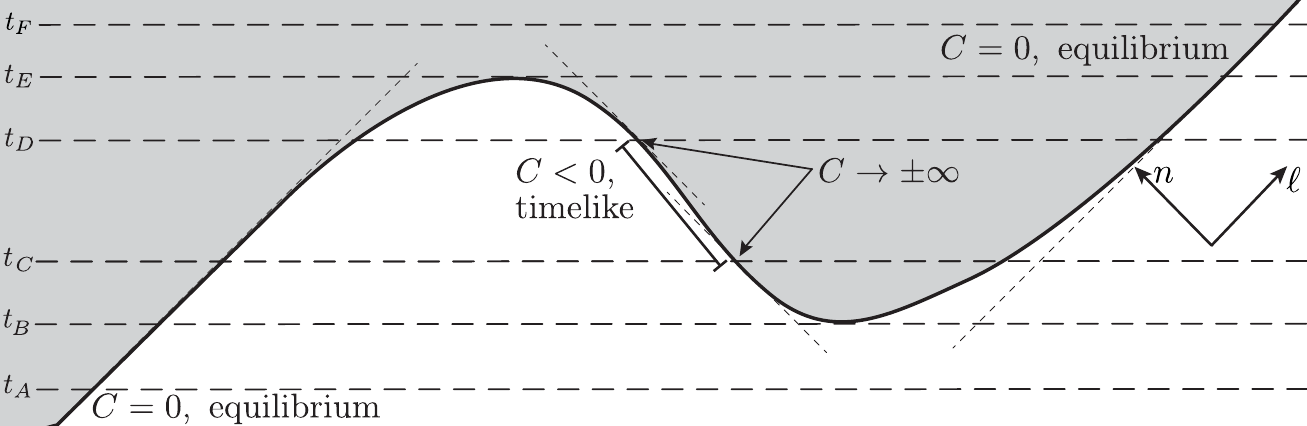}
\caption{Cartoon of a smoothly evolving MTT that exhibits an apparent horizon jump (adapted from \cite{Booth06}). 
The spacetime is foliated by a time parameter $t$. 
At $t_A$ there is a single isolated horizon in equilibrium with its surroundings. Matter then accretes and the MTT evolves as a spacelike 
dynamical horizon until $t_B$ where a dense concentration of matter outside the horizon causes a new marginally trapped 
surface to appear. If one was only tracking outermost marginally trapped surfaces 
then the apparent horizon would appear to jump at this
time. Relative to the time coordinate the new marginally trapped surface bifurcates into a pair of surfaces with one growing and one 
shrinking. At $t_C$ the inner surface becomes a timelike membrane and remains timelike until $t_D$ when it transitions back to being
spacelike. At both of these locations the MTT is parallel to $n$ and $C$ diverges.  At $t_E$ we return to a single outer horizon 
which relaxes to equilibrium at $t_F$. 
 }
\label{MTT_Figure}
\end{figure}
 An MTT demonstrating the full range of possible expansion parameters is shown in Figure \ref{MTT_Figure}.
There the MTT weaves back and forth through a time foliation so that  it may intersect a given instant multiple times. 
From the $(3+1)$ perspective there will then be 
multiple marginally trapped surfaces at some instants in time  (for example $t_C$ and $t_D$) and for a numerical code tracking only the 
outermost one as an apparent horizon, the horizon will appear to discontinuously jump (at $t_B$). If the MTT is spacelike, these 
multiple horizons may be viewed as resulting from foliation choices (see, for example \cite{Chu:2010yu}). However if the MTT has 
timelike or parallel-to-$n$ null sections there are unambiguous jumps: no foliation choice can remove them (examples may be found in 
\cite{BenDov:2004gh,Booth06})

$C$ depends on the scaling of the null vectors however we can also define an invariant measure of expansion.
Following \cite{Booth:2003ji,Booth07} we define an  \emph{evolution parameter} 
$\epsilon^2$ by 
\be
\frac{\epsilon^2}{r^2} \equiv  \frac{1}{2} C \tn^2 \, .
\ee
Despite the notation, $\epsilon^2$ will be negative if $C$ is negative\footnote{The parameter was 
originally developed for outermost slowly evolving horizons which are always null or spacelike if the energy conditions are satisfied. 
While we could take the absolute value of this quantity it will be convenient to retain the sign (which will tell us whether the MTT
is expanding or contracting).}. 
If $\cV$ is timelike or spacelike then with respect to $\hat{\cV}$, the unit normalized version of $\cV$,
\be
\left| \frac{1}{2} C \tn^2 \right| = \theta_{\hat{\cV}}^2 \, .
\ee
Thus $\epsilon^2$ is the square of the scaled and normalized rate of change of the area with a $\pm$ sign added to indicate whether $H$ is
spacelike or timelike.

This parameter vanishes in the isolated limit (even through $\hat{\cV}$ itself is
not well-defined there). When it is small the MTT is near equilibrium: 
a \emph{slowly evolving horizon} (see  \cite{Kavanagh:2006qe} for a range of examples). It diverges for null membranes which, as we
have seen, are associated with horizon jumps. Since these are about as far from equilibrium as one could imagine, this is fitting.

%  We will make much use of this parameter in later sections. 

\subsection{Marginally trapped tubes: dynamics}
\label{dynamics}

We can also consider MTT dynamics. By definition
\be
\Lie_{\cV} \tl = 0  \Longleftrightarrow  \Lie_\ell \tl - C \Lie_n \tl = 0 \, , \label{Ceq}
\ee
which for non-degenerate cases (we will return to degenerate cases where $\Lie_\ell \tl = \Lie_n \tl =  0$ in subsection \ref{ExEv})  implies that
\be
C = \frac{\Lie_\ell \tl}{\Lie_n \tl} %= \frac{G_{ab} \ell^a \ell^b}{\tilde{K} - G_{ab} \ell^a n^b} 
= \frac{T_{ab} \ell^a \ell^b}{1/(2 A)- T_{ab} \ell^a n^b} \, ,\label{C2}
\ee
where $A = 4 \pi r^2$. 
%We will return to degenerate cases where $\Lie_\ell \tl=\Lie_n \tl=0$ in Section \ref{ExEv}, but for now restrict our attention 
%to cases where this expression is well-defined.
From this expression we can understand how matter drives the evolution of MTTs. 

By the null energy condition the numerator is non-negative while  the sign of the denominator depends on the stability of the 
MOTS. %Momentarily leaving aside the degenerate $\Lie_n \tl = 0$
The classification is straightforward. 
First if $T_{ab} \ell^a \ell^b = 0$ then $C=0$ and we have an isolated horizon. Intuitively this makes sense.
 $T_{ab} \ell^a \ell^b$ is the flux across a surface with null normal $\ell_a$. In this case no matter crosses the horizon and so there is 
 no evolution. 

If $T_{ab} \ell^a \ell^b \geq 0$ then the qualitative behaviour of $C$ is determined by the relative size of the 
%$\Lie_n \tl$which in turn is decided by the relative size of the $(\ell, n)$ component of the
stress-energy tensor and the Gaussian curvature of the MOTS:
\begin{description}
\item[${1}/{(2A)}  > T_{ab} \ell^a n^b  \Rightarrow C>0$] $H$ is a dynamical horizon
\item[${1}/{(2A)} = T_{ab} \ell^a n^b  \Rightarrow C=\pm \infty$] $H$ is a null membrane 
\item[${1}/{(2A)} < T_{ab} \ell^a n^b  \Rightarrow C<0 $] $H$ is a timelike membrane \, . 
\end{description}
%
%Of course $\Lie_n \tl < 0 \Leftrightarrow 1/2A > T_{ab} \ell^a n^b$ is what one would generically expect: 
%in this case there are trapped surfaces inside the MTT while outside the $S(v,r)$ are untrapped. In the language
%of mathematical relativity such a horizon is \emph{stable} and this is a standard assumption used in many theorems
%about marginally outer trapped surfaces (for example \cite{Andersson:2005gq,Jaramillo:2011pg,Clement:2012vb}).
%
%
%
%
%\footnote{Comparing with 
%the mathematical relativity literature (for example \cite{Andersson:2005gq}), 
%when $\Lie_n \tl < 0$ then there are (many) outward pointing spacelike vectors $r^a$ for 
%which $\Lie_r \tl > 0$. With respect to spacetime foliations tangent to $r^a$, $S(v,r)$ is strictly stably outermost.}.
The first case is the standard one. Infalling matter drives the expansion of the horizon. 
The second and third are more exotic. There is a shrinking null or timelike membrane like those associated with horizon jumps
in Figure~\ref{MTT_Figure}. 

At least for  dust spacetimes (Oppenheimer-Snyder or Tolman Bondi) the physical origin of apparent horizon ``jumps'' is clear\cite{Booth06}. 
For timelike dust of density $\rho$ moving with four-velocity $u^a$ the stress-energy tensor is
\be
T^{\mbox{\tiny{TD}}}_{ab} = \rho u_a u_b \, . 
\ee
Then
\be
C_{\mbox{\tiny{TD}}} = \frac{1}{ \xi^2} \frac{2 \pi r^2 \rho}{1 - 4 \pi \rho r^2} \label{Cdust}
\ee
where $1/\xi =- 2 \ell^a u_a$ is a scaling parameter for the null vectors. 
%In those cases $G_{ab}\ell^a n^b$ is the dust density 
%$\rho = G_{ab} u^a u^b$, where $u^a$ is the unit tangent vector field to the streamlines and so 
The denominator is proportional to $\Lie_n \tl$ and 
\be
\Lie_n \tl \geq 0 \Longleftrightarrow \rho \geq \frac{1}{2A} \, .
\ee
Thus there is a dust density threshold set by the inverse horizon area beyond which the dust becomes dense enough to form 
a new black hole that contains the old one. As seen in FIG.~\ref{MTT_Figure}, a bubble of untrapped spacetime can remain 
caught between between the old and new apparent horizons but it quickly decays
as the timelike membrane approaches and annihilates the inner horizon.% (FIG.~\ref{MTT_Figure} again). 

%Given that the same condition $\Lie_n \tl =0$ defines extremality for isolated horizons and signals horizon jumps for dynamical ones it is 

%Thus if we are to consider evolutions from extremality we cannot ignore the degenerate cases of (\ref{Ceq}). In fact all of the 
%evolutions that we are interested in will begin in this state! In Sections \ref{exit} and \ref{ExEv} we will tackle this degeneracy. 
%However, given the double role of $\Lie_n \tl = 0$ in both horizon jumps and extremality it may not come as a surprise to see that some
%exits from extremality will be more dramatic than standard horizon evolutions. 
%

Though this is illustrative and useful in building intuition, timelike dust is not the focus of this paper. Instead we study null dust and (uncharged) 
scalar fields in a background electromagnetic field. 
First for (possibly charged) infalling null dust plus an electromagnetic field
\be
T^{\mbox{\tiny{NDEM}}}_{ab} = T^{\mbox{\tiny{ND}}}_{ab}  + T^{\mbox{\tiny{EM}}}_{ab} =  \mu n_a n_b + T^{\mbox{\tiny{EM}}}_{ab} \, , 
\ee
so that
\be
C_{\mbox{\tiny{NDEM}}} = \frac{8 \pi  \mu r^4}{ r^2 - q^2} \, . \label{CVad}
\ee
For charged dust $q$ will be dynamical but for uncharged dust it is constant. Then if $r>|q|$ the expansion will be 
positive and spacelike while inside $r<|q|$ it will be negative and timelike. We will see this behaviour for outer and inner horizons.

%
%
%and the extremality bound $r\leq q$ 
%so for strictly stably outermost MTTs only isolated or dynamical horizons are possible. Note however that extremal MTTs and those with trapped surfaces outside
%are not strictly stably outermost. Thus an extremal MTT can be a null membrane and a inner MTT can be either isolated or a timelike membrane. 

%
%By the energy conditions $\mu \geq 0$ and so it is clear that with these matter fields null membranes will only be possible at extremality $r=q$. 

The second type of matter consists of an uncharged  scalar field obeying the Klein-Gordon equation
\be
\nabla^2 \varphi = m_o^2 \varphi \, ,
\ee
(where $m_o$ is the mass of the field) along with an electromagnetic field so that
\be
T^{\mbox{\tiny{KGEM}}}_{ab} =  \frac{1}{4 \pi} \left(\nabla_a \varphi \nabla_b \varphi - \frac{1}{2} (\nabla_c \phi \nabla^c \varphi+  m_o^2 \varphi^2) g_{ab} \right) + T^{\mbox{\tiny{EM}}}_{ab} \, . 
\ee
Thus the expansion parameter  is
\be
C_{\mbox{\tiny{KGEM}}} =\frac{2r^4 (\Lie_\ell \varphi)^2}{r^2 - q^2-m_o^2 \varphi^2 r^4} \, , \label{Cscalar}
\ee
where $q$ is constant since there is no means of propagating charge in these spacetimes. For a massless scalar field the qualitative evolution is again determined by whether or 
not a MOTS is outside or insider $r=q$. However for a massive scalar field more complicated evolutions like those in FIG.~\ref{MTT_Figure} may be possible\footnote{ 
As far as we know explicit timelike membrane examples have not yet been constructed. An attempt to construct such examples in \cite{Booth06} was not successful though 
the difficulties obstructing the construction were probably numerical rather than physical.}. 

%
%
% As usual the strictly stably outermost condition is equivalent to 
%the denominator being positive. Thus 
%
%
% For a massless scalar field the strictly stably outermost condition requires $r\geq q$ and so 
%we have the same restrictions on horizon behaviours as for null dust. For a massive scalar field things are not 
%
%
%
% By the extremality bound we have the same restrictions on horizon 
%behaviours as for null dust. 

%Finally we will briefly consider a massive uncharged scalar field (with no electromagnetic field). Then 
%\be
%\nabla^2 \varphi = m_o^2 \varphi
%\ee
%and
%\be
%T_{ab}^{\mbox{\tiny{MKG}}} =  \frac{1}{4 \pi} \left(\nabla_a \varphi \nabla_b \varphi - \frac{1}{2} (\nabla_c \nabla^c \varphi +m_o^2 \varphi^2 ) g_{ab} \right) 
%\ee
%so 
%\be
%C_{\mbox{\tiny{MKG}}} = \frac{2r^2 (\Lie_\ell \varphi)^2}{1 - 2 r^2 m^2 \varphi^2} \, .  \label{CMKG}
%\ee
%In this case more general behaviours like those of timelike dust depicted in FIG.~\ref{MTT_Figure} are possible. 

\section{Evolutions for null dust}
\label{nulldust}

We now consider our most detailed example. Vaidya Reissner-Nordstr\"om spacetimes contain infalling, possibly charged, null dust. In
this section we introduce these spacetimes and then use them to study possible exits from extremality along with dynamical extremal horizons. 

%
%We start with the
%case where non-extremal matter accretes onto the MTT and so it immediately becomes non-extremal. In this case we can evaluate
%(\ref{epsilonpm}) as a limit. However we will also consider the case where a horizon remains extremal throughout its evolution. In that case 
%the dynamical approach fails and we have to rely on kinematics. 

\subsection{Vaidya Reissner-Nordstr\"om and its horizons}
\label{VRN}

\subsubsection{The spacetime}
In Eddington-Finkelstein-like coordinates infalling Vaidya Reissner-Nordstr\"om (VRN) spacetimes \cite{Bonnor:1970zz}
are described by
\be
ds^2 = - \left( 1 - \frac{2 m(v)}{r} + \frac{q(v)^2}{r^2} \right) dv^2 + 2 dv dr + r^2 d\Omega^2 
\ee
where $m(v)$ and $q(v)$ respectively determine the rate of accretion of mass and charge. The electromagnetic field
is defined by the potential 
\be
A = -\frac{q}{r} dv \label{potential}
\ee
and the full dust plus electromagnetic stress-energy tensor is
\be
%T_{ab} = \left( \mu\subs{d} + \mu\subs{q} \right)  [dv]_a \otimes [dv]_b + T_{ab}^{\mbox{\tiny{EM}}} \, ,
T_{ab} = \mu  [dv]_a \otimes [dv]_b + T_{ab}^{\mbox{\tiny{EM}}} \, ,
\ee
%where we have broken the dust energy density into two pieces
%\be
%\mu\subs{d} =\frac{\dot{m}}{4 \pi r^2}   \; \; \mbox{and}  \; \; 
%\mu\subs{q}  = - \left( \frac{q}{r} \right) \frac{ \dot{q}}{4 \pi r^2}  \label{muq}
%\ee
%which are roughly the energy associated with the the mass of the dust and that associated with the charge (note that this looks like a charge density
%times an electromagnetic potential). Later on we will find it convenient to have these separated. 
where 
\be
\mu = \frac{r \dot{m} - q \dot{q}}{4 \pi r^3} \label{mu}
\ee  
is the energy density of the dust\footnote{Rewriting as 
\be
\mu = \frac{\dot{m}}{4 \pi r^2} - \left( \frac{q}{r} \right) \frac{ \dot{q}}{4 \pi r^2} 
\ee
this can be seen to have two components: one associated with the dust mass energy and the other with the potential energy of the 
dust relative to the electromagnetic field. \label{Foot5}}. 
The last part, $T_{ab}^{\mbox{\tiny{EM}}}$ is the electromagnetic field stress-energy
that is generated by $A$. 

Note that this stress-energy tensor is consistent with the Maxwell equations which tell us that the electromagnetic field is supported by a (null) current
\be
j_{\mbox{\tiny{EM}}}^b = -\frac{1}{4 \pi}  \nabla^a F_a^{\phantom{a}b} =   -\frac{\dot{q}}{4 \pi r^2} \left( \frac{\partial}{\partial r}\right)^b \label{EMflow}\,. 
\ee

The spacetime satisfies all the standard energy conditions if and only if
\be
\dot{m} r - \dot q q  \geq 0 \, . \label{EC}
\ee
In particular note that if $q\dot{q} = 0$ then the energy conditions are satisfied as long as $\dot{m} \geq 0$ and so the horizon is expanding as it 
absorbs positive energy density dust. If $q \dot q\neq 0$  then there will always be energy condition violations for
\be
r < \left( \frac{\dot{q}} {\dot{m}}\right) q \, . 
\ee
%If $\dot{q} > \dot{m}$ these may be outside the 
%
%
%In the following we assume that $\dot{q} \leq \dot{m}$ so that any energy condition violations will be cloaked inside the horizon. Since we will chiefly be 
%interested in matter accreting onto (at least initially) extremal $q=m$ horizons this will also mean that we do not consider matter that can super
%
%
%otherwise the infalling dust would be super-extremal with a charge density higher than its mass density and could 
%super-charge the black hole). Thus  any such region will necessarily be clothed if there is a strictly stably outermost MTT. 
 We will return to these apparent violations  in the next subsection and then again in \ref{ExEv}.

%Note that for an initially extremal horizon ($r_H = q$) such violations will occur outside the horizon for $\dot{q} > \dot{m}$ but otherwise be confined inside. Not 
%coincidentally for $\dot{q} > \dot{m}$ that initially extremal horizon would immediately evolve to become a super-extremal naked singularity:
%the energy condition violations correspond to the charged dust continuing to move inwards even though the electrical repulsion has become 
%stronger than the gravitational attraction.

\subsubsection{Marginally trapped tubes}

As a first step to locating horizons, we define:
\bea
\ell &=& \frac{\partial}{\partial v} + \frac{1}{2} \left(1 - \frac{2m}{r} + \frac{q^2}{r^2} \right) \frac{\partial}{\partial r} \; \;  \mbox{and} \\
n &=& - \frac{\partial}{\partial r} \, . 
\eea
Then the geometrical quantities associated with these null vector fields are the expansions
\bea
\tl &=& \frac{1}{r} \left( 1 - \frac{2m}{r} + \frac{q^2}{r^2} \right)  \; \; \mbox{and}  \label{tLVRN}\\ 
\tn &=& - \frac{2}{r} \,  , \label{tn}
\eea
and the inaffinities
\bea
\kappa_\ell &=& \frac{rm-q^2}{r^3} \; \; \; \mbox{and} \; \; \;  \kappa_n =0  \label{kappal} \; ,
\eea
while the non-zero components of the stress-energy tensor are
\bea
T_{ab} \ell^a \ell^b & = &  \mu  \; \; \mbox{and} \label{TinLL}\\
T_{ab} \ell^a n^b & = &  \frac{q^2}{8 \pi r^4}  \label{TinLN} \, . 
\eea

Thus by (\ref{tLVRN})  there are MTTs at 
\be
r_{\pm}(v) = m \pm \sqrt{ m^2 - q^2} \, , \label{rpm}
\ee
where $r_+$ and $r_-$ are the inner and outer horizons. This is exactly the same relation as for regular RN. 

Knowing the location of the outer MTT provides some extra insight into the noted violations of the energy condition.
Focussing on an extremal horizon 
($r_H = m=  q$) such violations will occur outside the horizon when $\dot{q} > \dot{m}$ but otherwise are confined inside. 
However if $\dot{q} > \dot{m}$ 
it is also true that an initially extremal horizon would immediately evolve to become a super-extremal naked singularity.

The energy condition violations correspond to the charged dust continuing to move inwards even though the electrical repulsion has become 
stronger than the gravitational attraction. This last conclusion continues to hold more generally: energy condition violations signal 
unphysical dust evolution. In Section \ref{ExEv} we will examine how this may be corrected but for now assume $\dot{q} < \dot{m}$ to avoid 
this difficulty for our extremal horizons. 

Next consider dynamics of the MTT. From  (\ref{CVad}) the expansion parameter of these MTTs is
%
%Next at general $r$,
%\bea
%G_{ab} \ell^a \ell^b &=& \frac{2}{r^3} \left(\dot{m} r - q \dot{q}  \right) \; \mbox{and} \label{Gll} \\
%G_{ab} \ell^a n^b & = & \frac{q^2}{r^4} \, 
%\eea
%so from (\ref{Lltl}) and (\ref{Lntl})
%\bea
%\Lie_\ell \tl &=& -  \frac{2}{r^3} \left(\dot{m} r - q \dot{q} \right) \; \mbox{and} \\
%\Lie_n \tl & = & - \frac{1}{r^4} \left( r^2-q^2 \right) \, .
%\eea
%Extremality and/or horizon jumps occur when $|q| = m$. 
%
%Substituting the above quantities into  (\ref{C2}) the expansion parameter on (non-degenerate) inner and outer horizons  is
\be
C_\pm = \pm   \left( \frac{r_\pm \dot{m} - q \dot{q}}{\sqrt{m^2  - q^2}} \right) \,, \label{Cpm}
\ee
where one consistently chooses the positive (minus) signs to get the outer (inner) horizon. By
(\ref{EC}), $C_+ \geq 0$ and $C_- \leq 0$: the outer horizon is spacelike (or null) and non-contracting while the inner one is 
timelike (or null) and non-expanding. For $|q| = m$ this diverges if the numerator is non-zero or is ill-defined if it vanishes. 
That said, if it  immediately exits extremality then we can (and will)  study the departure as a limiting process.

The corresponding (scaling invariant) evolution parameter is
\be
\epsilon^2_\pm =   2   \left( \dot{m}  \pm \frac{m \dot{m}- q \dot{q}}{\sqrt{m^2 - q^2}} \right) \, \label{epsilonpm}
\ee
and so this invariant parameter also diverges or is ill-defined for extremal horizons. 
By (\ref{rpm}) it can be written in an 
even simpler form:
\be
\epsilon^2_\pm  = 2 \dot r_{\pm}  \label{dotr}
\ee
and so discontinuities in the expansion parameter are equivalent to discontinuities in the rate of change of the areal radius.

\subsection{Exit from equilibrium}
\label{exit}

Now that we have a concrete (and exactly solvable) model we can study horizon evolutions. We begin with the accretion
of uncharged dust onto both extremal and (for comparison) nearly extremal MTTs. 

\subsubsection{Possible exits}
Take a black hole with initial charge $q_o$ and mass $m_o \geq |q_o|$ and without loss of generality assume the accreting dust first crosses the horizon at
$v=0$ (so  $m=m_o$ and $q=q_o$ for $v \leq 0$).
We are interested in the transition from equilibrium and so expand $m(v)$ as a Taylor series and 
consider only the leading term. Thus for a time scale $v_o$ and $0 \leq v \ll v_o$:
\be
m(v) = m_o\left(1+  (v/v_o)^k + O \left( \frac{v}{v_o}\right)^{k+1} \right)
\ee
for some positive integer $k$. 

For such a mass function the evolution parameter takes different forms for  non-extremal versus initially extremal horizons. For $0 \leq v \ll v_o$,  (\ref{dotr}) gives:
\be
\epsilon_\pm^2 \approx \left\{ 
\begin{array}{ll}
\frac{2km}{v_o} \left( 1 \pm \frac{1}{\sqrt{1 - (q_o/m_o)^2}} \right) \left( \frac{v}{v_o} \right)^{k-1}   & q_o < m_o  \\
\pm \left( \frac{\sqrt{2}km_o}{ v_o} \right) \left( \frac{v}{v_o} \right)^{k/2 -1} & q_o = m_o  \, 
\end{array}
\right.  
\ee
so the details of transitions are determined by the limits of these expressions as $v \rightarrow 0$.

\begin{figure}
\includegraphics[scale=0.9]{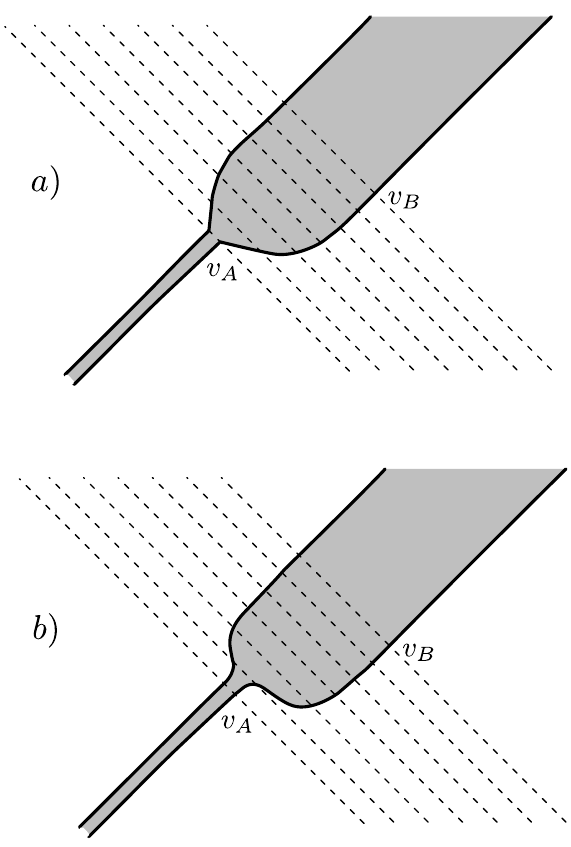}
\caption{Evolution of an initially non-extremal RN horizon. The trapped region is shaded  and 
null directions are at $45^\circ$ to the horizontal as in FIG. \ref{MTT_Figure}. The accreting dust begins crossing the horizon
at $v=v_A$ and finishes at $v_B$. It is represented by dashed lines. a) represents the $k=1$ case where the rate of expansion jumps discontinuously
to a finite value as the first matter arrives while for b) $k>1$ the rate is continuous. In both cases the inner MTT becomes timelike 
while the outer MTT becomes spacelike. }
\label{NExExit}
\end{figure}
Focussing first on the non-extremal horizon there are two classes of transition from isolation:
\be
\lim_{v \rightarrow 0} \left.  \epsilon_\pm^2 \right|_{\mbox{\tiny{non ex}}} =
\left\{
\begin{array}{ll}
\frac{2m}{v_o} \left( 1 \pm \frac{1}{\sqrt{1 - (q_o/m_o)^2}} \right) & k = 1 \\
0  & k>1
\end{array} 
\right. \, . 
\ee
For $k>1$ the evolution parameter changes continuously as the matter arrives but for $k=1$ (a linear increase in mass) the evolution parameter jumps
discontinuously. Instantaneously the inner and outer horizons become a timelike membrane and a dynamical horizon. However, this discontinuity is not really 
a surprise since by (\ref{mu}) there is also a discontinuity in the Ricci tensor\footnote{
This discontinuity is physically caused by the jump of the dust energy 
density from zero to a finite value at $v=0$. It is not caused by the presence of a  thin shell discontinuity (null or otherwise). By the junction 
conditions for null surfaces (\cite{clarke1987junction,Barrabes:1991ng,Mars:1993mj} or \cite{poisson2004relativist} for a textbook presentation) the energy density of any 
such shell at $v=0$ would be proportional to the jump in the inner expansion $\theta_{(n)}$. This is zero by (\ref{tn}). The pressure in such a shell would be proportional to the 
jump in $\kappa_\ell$. This is also zero by (\ref{kappal}).}. The two possibilities are depicted schematically in FIG.~\ref{NExExit}. 
%
%These correspond to our observations for the $\mu_1$ versus $\mu_{>1}$ cases. Both classes are depicted 
%schematically in FIG.~\ref{NExExit} where the jump in the $\epsilon^2$ for $k=1$ shows up as a kink in 
%both the inner and outer MTTs. Conversely for $k>1$ the continuous expansion parameter corresponds to smoother MTTs. 

% However for 
%$k>1$ as the first matter impacts the horizon the rate of expansion is zero and then it gradually increases. This smoother evolution is shown
%in FIG. \ref{Exit}a). 
%Again this is not 
%surprising since in this case $\Lie_n \tl >0$ and so the denominator of  (\ref{C2}) is finite. 

\begin{figure}
\includegraphics[scale=0.9]{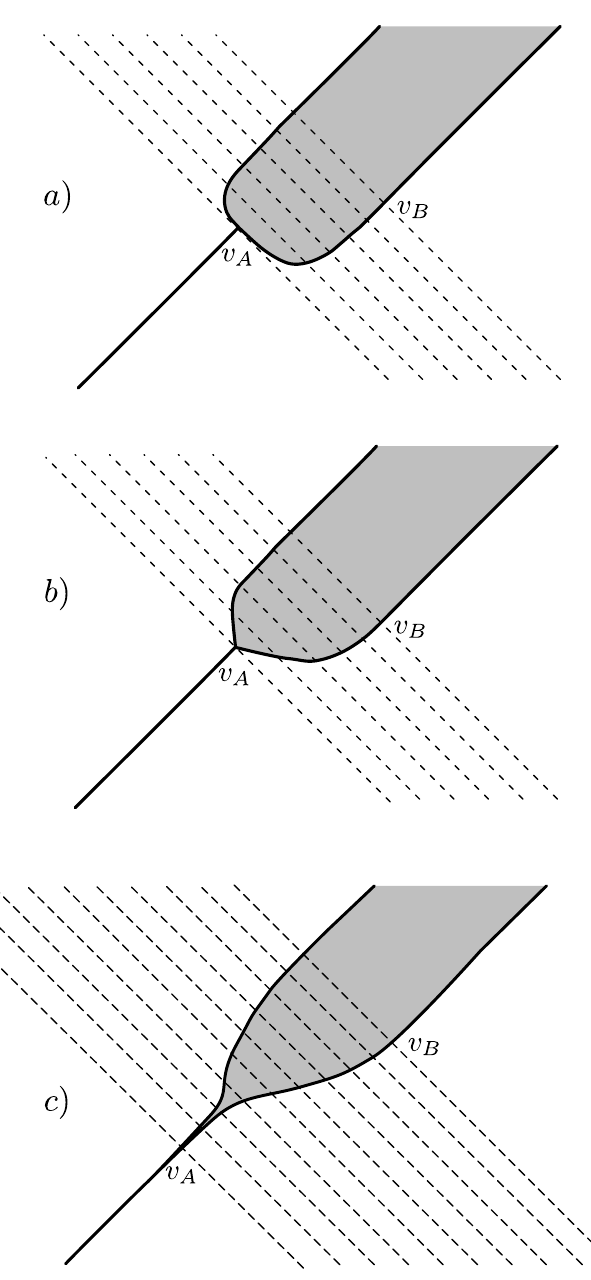}
\caption{Evolution of an initially extremal RN horizon. Diagrams are set up in the same way as in FIG. \ref{NExExit}. Respectively they are
a) $k=1$, b) $k=2$ and c) $k>2$. }
\label{ExExit}
\end{figure}
For an initially extremal horizon things are more interesting. The evolution parameter is
\be
\lim_{v \rightarrow 0} \left. 
\epsilon_\pm^2 \right|_{\mbox{\tiny{ex}}} 
=  \left\{
\begin{array}{ll}
\pm \infty  & k=1 \\
\pm \frac{2\sqrt{2}m_o}{ v_o} & k = 2 \\
0  & k \geq 3
\end{array} 
\right.  \, .
\ee 
This time there are three characteristic evolutions from extremality which are depicted schematically in FIG.~\ref{ExExit}. 
For $k=1$ the arrival of matter causes the MTT to jump from being isolated and parallel to $\ell$ to being maximally evolving and parallel to $n$. 
For $k=2$ the situation is analogous to the $k=1$ case for non-extremal horizons: the jump is discontinuous to a timelike inner and spacelike outer MTT. 
However in this case the Ricci tensor is continuous and the discontinuity instead follows from the extremality. 
Finally for $k\geq3$, the evolution parameter  is continuous. 
%
%There are discontinuities in $\epsilon^2$ for both $k=1$ and $k=2$ of which $\mu_1$ and $\mu_2$ are examples. 
%As noted for the earlier examples the  $\infty$ for $k=1$ isn't quite as serious as it might
%first appear: it signals that the MTT becomes parallel to $n$ (surfaces of constant $v$). So the difference between the $k=1$ 
%and $k=2$ cases is more of degree rather than kind. That said it is
%still a dramatic behaviour: a discontinuous jump from equilibrium to a maximal rate of evolution. Finally for $k>2$ the initially extremal 
%horizon smoothly bifurcates into an inner timelike membrane and an outer dynamical horizon. All possibilities are  
%shown schematically in FIG. \ref{ExExit}.

Regardless of the details, in all three of these cases we have a tripartite MTT: the initially extreme horizon splits into two non-extremal pieces 
which remain distinct. However for the extremal $k=1$ case we can also recognize a more familiar
situation if we temporarily disregard the initial state: the timelike membrane and dynamical horizon connect at 
an instantaneous null membrane.
 This is similar to the  behaviour observed during horizon jumps.% like FIG. \ref{MTT_Figure}. 

\subsubsection{Complete evolutions}

%
%
%\textit{To begin note from (\ref{C2}) that if $\Lie_n \tl = 0$ while $\Lie_\ell \tl \neq 0$
%then $C$ diverges and from our experience in the previous section we expect this 
%to mean that the MTT is tangent to $n$. However if we are considering the evolution of an initially extremal MTT as matter accretes onto
%it then things are a bit more complicated: the numerator will also initially be zero and so we must consider the limiting process in 
%the departure from equilibrium. To explore the possibilities we consider a concrete model. 
%}

%\begin{figure}
%\includegraphics[scale=0.8]{DiscontExit.pdf}
%\caption{Schematic of possible evolutions of Vaidya Reissner-Nordstr\"om horizons. The trapped region is shaded  and 
%null directions are at $45^\circ$ to the horizontal as in FIG. \ref{MTT_Figure}. The accreting dust begins crossing the horizon
%at $v=v_A$ and finishes at $v_B$. It is represented by dashed lines. In a) the dust accretes onto a non-extremal horizon 
%while in the other cases the horizon is initially extremal. For b) $m \propto v$ at leading order and so the slope of the MTT is
%discontinuous jumping from being parallel to $\ell$ to parallel to $n$. However for c) $m \propto v^k$ with $k > 1$ and the horizon
%instead bifurcates in a less dramatic way. 
% }
%\label{Exit}
%\end{figure}

Informed by these observations we can also consider full evolutions which include a return to equilibrium. As before we fix 
$q=q_o$ and now consider piecewise mass functions of the form
\be
m(v) = \left\{
\begin{array}{ll}
 m_o & v<0\\
 m_o (1 + m_\Delta \mu(v)) & 0 \leq v \leq 1 \\
 m_o (1 + m_\Delta) & v>1
\end{array}
 \right. \, ,
\ee
where $\mu$ is a continuous function with $\mu(0) = 0$ and $\mu(1)=1$. Thus $m(v)$ is similarly continuous though not
necessarily differentiable. 
For such mass functions the initial and final areal radii of the inner and outer horizon are respectively
\bea
r_o^{\pm} &=& m_o \pm \sqrt{m_o^2 - q_o^2} \; \; \; \mbox{and} \\
r_f^{\pm}  &=& (m_o +m_\Delta) \pm \sqrt{(m_o + m_\Delta)^2 - q_o^2} \, . 
\eea

Our examples take near-extremal $q_o = 0.999 m_o$ and extremal $q_o = m_o$ horizons as initial states. We take
$m_\Delta = {m_o}/{20}$ and consider three increasingly smooth $\mu(v)$:
\bea
\mu_1 &=& v \\
\mu_2 &=& -v^2(2v-3) \nn\\
\mu_3 &=& v^3(6v^2-15v+10) \nn
%\mu_4 &=& -v^4(20v^3-70v^2+84v-35) \nn \\
%\mu_5 &=&  v^5(70v^4-315v^3+540v^2-420v+126) \nn \; . 
\eea
The subscript indicates at which order of derivative the full $m$ becomes discontinuous. Thus for $\mu_1$, the first derivative of the
mass function is discontinuous at $v=0$ and $v=1$, while for $\mu_3$ the discontinuity doesn't show up until the third order derivative. 
%For the initially extremal case we evaluate $C(0)$ by taking the limit of (\ref{Ceq}) as $v \rightarrow 0^+$. 

\begin{figure}
\includegraphics{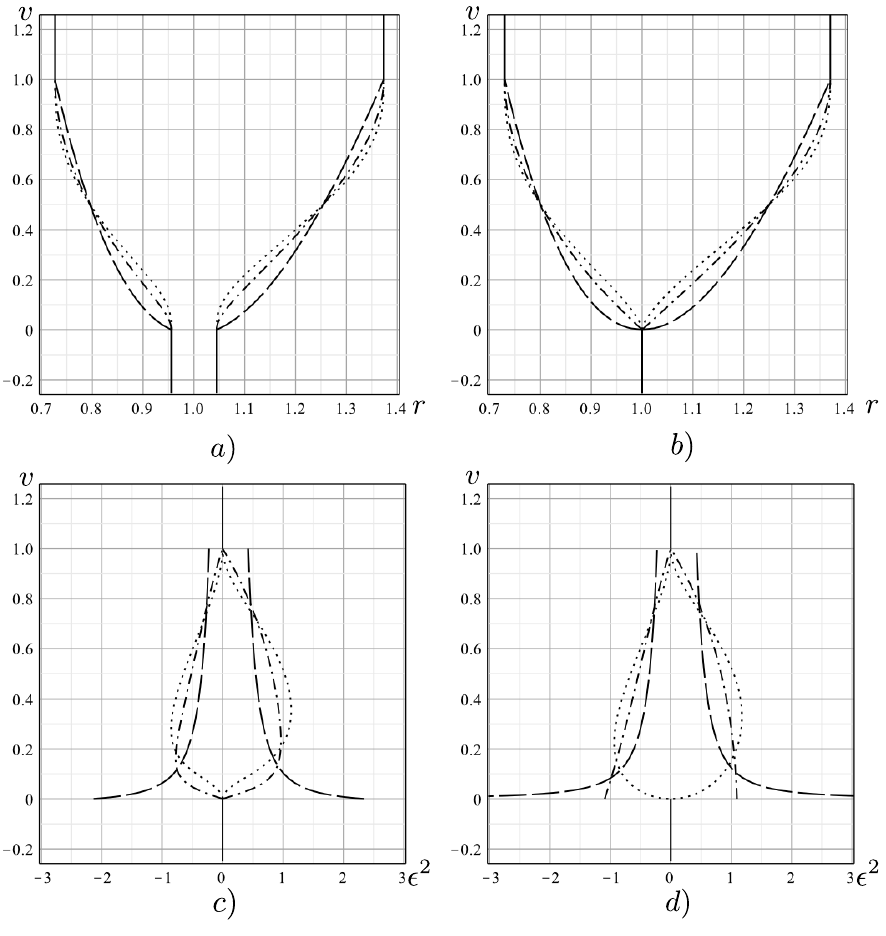}
\caption{Inner and outer horizon evolutions for a) near extremal ($q_o=0.999m_o$) and b) extremal ($q_o=m_o$) VRN spacetimes. 
In both cases the areal radius $r$ is on the horizontal axis and Eddington-Finkelstein time $v$ on the vertical (both in units of $m_o$). 
Below these subfigures, c) and d) plot the corresponding evolution parameters. 
In all four subfigures the different curves represent accreting matter of varying degrees of smoothness. Long dashes are $\mu_1$, 
dot-dashes $\mu_2$ and dots $\mu_3$. When all models behave in the same way 
a solid black line is used. }
\label{MapleEv}
\end{figure}
These MTTs and the corresponding expansion parameters are plotted in FIG. \ref{MapleEv}. The expected discontinuities from the 
earlier analysis are seen: for the initially extremal case there is an infinite discontinuity for $k=1$ and finite for $k=2$ while for 
non-extremal there is only a finite discontinuity for $k=1$. At the other end of the evolution ($v=1$) there are discontinuities in the 
evolution parameter for $k=1$ (again reflecting the corresponding discontinuity in the Ricci tensor) while for higher $k$ it 
returns to equilibrium continuously.

\subsection{Evolution at extremality}
\label{ExEv}

With the accretion of matter the MTTs of the last section immediately became non-extremal. In this subsection we 
consider another case: the MTTs remain extremal throughout their evolution. Such evolutions are generated by 
$q(v)=m(v)$ dust accreting onto an extremal horizon. With this matter, the energy conditions require
\be
\dot{m} r - \dot{q} q = \dot{m} (r - m)  \geq 0 \; \; \mbox{for} \; r \geq m \, . 
\ee
The horizon of an extremal VRN is at $r_{\mbox{\tiny{ex}}} = m$ and so, as shown in FIG.~\ref{PreMerge}a) the energy conditions hold on and outside the horizon. 
What happens inside is a bit more complicated. 

For VRN spacetimes where the corresponding violations occur outside the horizon it has been argued by Ori \cite{Ori1991} 
that they are  indicative of regions where the solution is no longer physically meaningful. They occur in regions where 
the electromagnetic repulsion has become stronger than the gravitational attraction yet the solution still requires the dust to move inwards. 
Ori resolved this physical inconsistency by surgically removing the problematic region and replacing it with an outgoing VRN spacetime matched  across the transitional region. Then the newly constructed solution described a shell that fell inwards until the electromagnetic repulsion 
caused it to bounce back outwards. In that paper careful physical arguments are used to motivate this construction. 

In this section we will implement a similar resolution for extremal VRN. In preparation for this we review outgoing VRN. 

\subsubsection{Outgoing VRN}
\begin{figure}[h]
\includegraphics[scale=0.9]{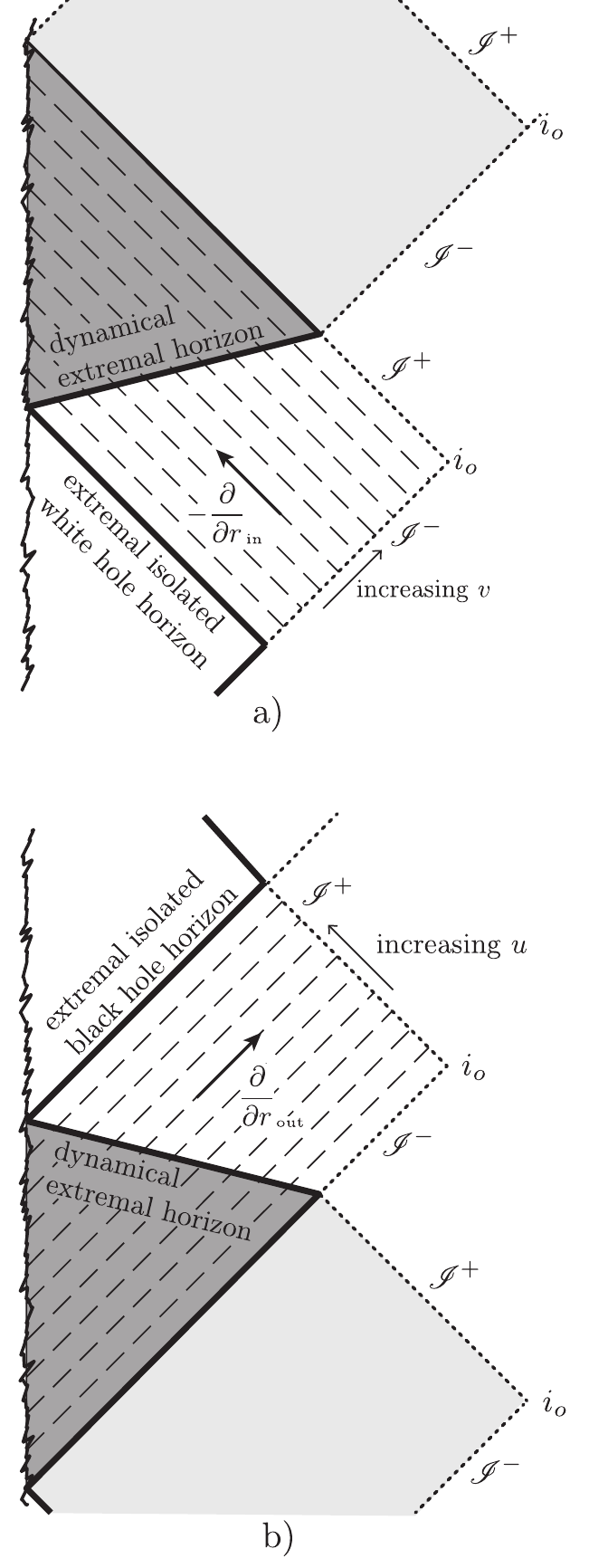}
\caption{Extremal dynamical Vaidya spacetimes. In a) infalling Vaidya,  $\dot{q} = \dot{m}$ dust falls through the black hole horizon causing it to expand as a spacelike surface. 
In b) outgoing Vaidya, $\dot{q} = \dot{m}$ dust is emitted from a white hole horizon causing it to shrink, again as a spacelike surface. In both cases regions that will be excised in 
the construction of FIG.~\ref{DynEx} are shaded in (light or dark) gray. The region where the energy conditions are violated is shaded darker and there the dust continues to move inwards (or outwards) even though the electromagnetic force acting on the dust is stronger than the gravitational one.   }
\label{PreMerge}
\end{figure}

Just as the ingoing VRN metric closely resembles Reissner-Nordstr\"om in ingoing Eddington-Finkelstein coordinates,
so does the outgoing VRN metric resemble Reissner-Nordstr\"om in outgoing Eddington-Finkelstein coordinates:
\be
ds^2 =- \left( 1 - \frac{2 M(u)}{r_{\subs{out}}} + \frac{Q(u)^2}{r\subs{out}^2} \right) du^2 - 2 du dr_{\subs{out}}+ r_{\subs{out}}^2 d \Omega \, .
\ee
This time $u$ labels the outgoing radial null geodesics. $r\subs{out}$ is again the areal radius of spherical shells however we add the subscript to emphasize that $\partial/\partial r\subs{out}$ is a very different vector than it is for ingoing Vaidya (in particular it is future-outward rather than 
past-outward pointing). As is the case for RN in these coordinates, this version of VRN describes
a white hole.

The spacetime satisfies all of the energy conditions if and only if
\be
Q Q'- r\subs{out}M' \geq 0 \, , 
\ee
where derivatives with respect to $u$ are indicated by primes. Note that if $QQ'=0$ we must have $M$ non-increasing and so the white hole shrinks as it emits positive
energy dust.
 If $QQ' \neq 0$ and $M'<0$ then there will always be energy condition violations for 
\be
r\subs{out} < \left| \frac{Q'}{M'} \right| Q \, .
\ee
This is all depicted (for the extremal case) in  FIG.~\ref{PreMerge}b).

Turning to geometry and horizons, a suitable pair of outward and inward future null vectors are 
\bea
\ell &=& \frac{\partial}{\partial r\subs{out}} \; \; \mbox{and} \\
n &=& \frac{\partial}{\partial u} 
- \frac{1}{2} \left(  1 - \frac{2 M}{r\subs{out}} + \frac{Q^2}{r\subs{out}^2} \right) \frac{\partial}{\partial r\subs{out}} \,  ,
\eea
which have expansions
\bea
\theta_{(\ell)} &=& \frac{2}{r\subs{out}} \; \; \mbox{and} \\
\theta_{(n)} & = & - \frac{1}{r\subs{out}} \left(1- \frac{2M}{r\subs{out}} +\frac{Q^2}{r\subs{out}^2} \right) \, . 
\eea
In this case the \emph{inward} expansion vanishes at 
\be
r_{\pm}{\subs{out}}(u) = M(u) \pm \sqrt{ M(u)^2 - Q(u)^2} \, , \label{rpm2}
\ee
and between these roots spacetime is anti-trapped (both expansions are positive). Of course for an extremal RN solution, this 
anti-trapped region vanishes. 

The non-zero components of the stress-energy tensor also take a similar form to that for infalling VRN:
\bea
T_{ab}  \ell^a n^b & = &  \frac{Q^2}{8 \pi R^4} \; \mbox{and} \label{ToutLN} \\
T_{ab}  n^a n^b & = & \frac{Q Q'-RM'}{4 \pi R^3} \, .  \label{ToutNN}
\eea

\subsubsection{Matching infalling to radiating extremal VRN}

\begin{figure}
\includegraphics[scale=0.9]{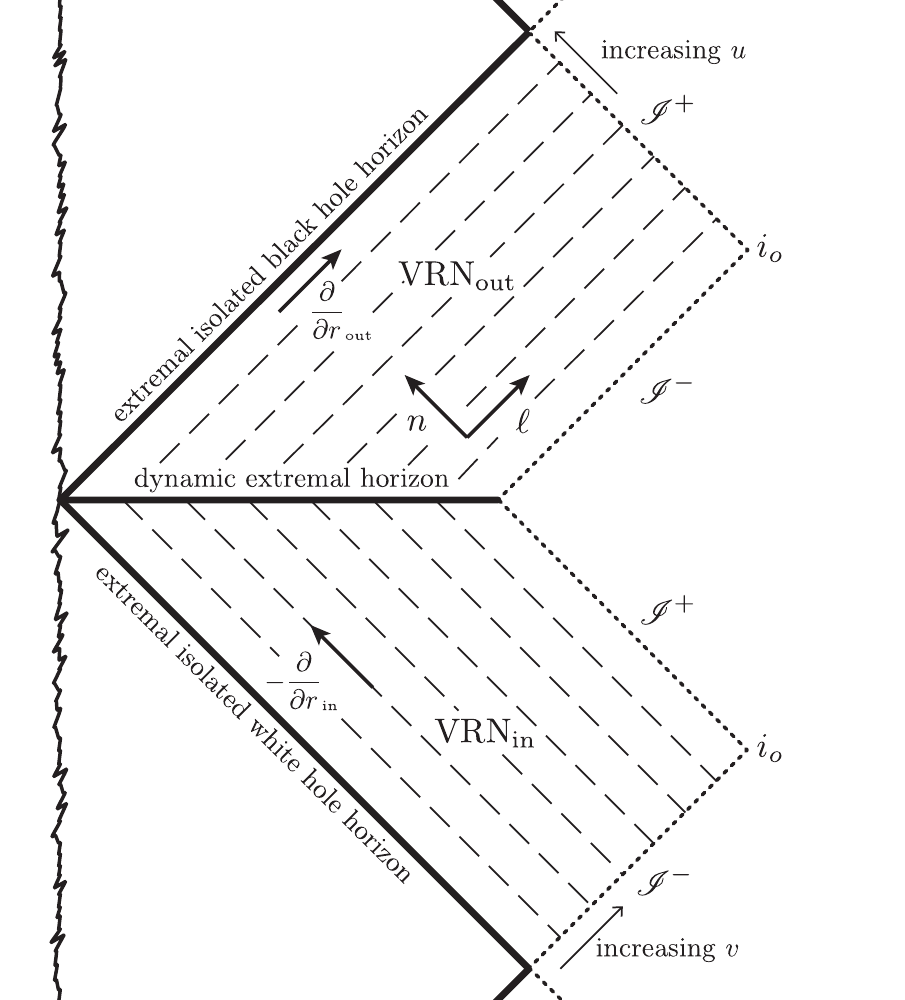}
\caption{A Carter-Penrose diagram of an extremal RN black hole irradiated with extremal null dust. 
$\mbox{VRN}_{\mbox{\tiny{in}}}$ is a region of 
ingoing VRN where dust accretes onto the black hole while $\mbox{VRN}_{\mbox{\tiny{out}}}$ is a region of outgoing VRN where
a white hole radiates dust. They meet on a common MTT which is dynamical and spacelike.  
The jagged line represents the timelike singularity at $r=0$ 
while the dotted outer lines are future and past null infinities. 
 Note that there  are no trapped regions.  }
\label{DynEx}
\end{figure}

We are now ready to construct our dynamical extremal VRN spacetime. Respectively we set $q=m$ and $Q=M$ and assume that 
that $\dot{m}>0$ and $M'<0$ at all times so that the MTTs at $r\subs{in}=m$ and $r\subs{out}=M$ are always spacelike (this simplifies the construction). Then 
as shown in FIG.~\ref{PreMerge} for $r\subs{in}<m$ and $r\subs{out}<M$  the energy conditions are violated in both spacetimes. We excise those regions and connect the remaining sections of the spacetime along the extremal MTT as shown in FIG.~{\ref{DynEx}. We can 
analyze this connection by checking the usual junction conditions: in this case for spacelike surfaces\cite{Israel66,clarke1987junction,Barrabes:1991ng,Mars:1993mj,poisson2004relativist}. 

Starting with the infalling (black hole) side of the geometry
we parameterize the MTT with induced coordinates $y\subs{in}^i=\{v,\theta,\phi\}$ (note the use of mid-alphabet latin indices to indicate 
we are referring to  the MTT). Then the MTT three-metric is 
\be
h\sups{in}_{ij} dy\subs{in}^i dy\subs{in}^i = 2\dot{m}   dv^2 + m^2  d \Omega^2 \, . \label{hin}
\ee
As expected for $\dot{m} \geq 0$ this is spacelike. Correspondingly for the radiating (white hole) side we parameterize the MTT with induced coordinates 
$y\subs{out}^i=\{u,\theta,\phi\}$ and find the three-metric  is
\be
h\sups{out}_{ij} dy\subs{out}^i dy\subs{out}^i   =  - 2 M' du^2 + M^2 d \Omega^2 \, . \label{hout}
\ee
For $M'<0$ this is spacelike. 

We can match the ingoing and outgoing spacetimes induced metrics on the MTT, with the (on MTT) coordinate transformation  
\be
u=-v \label{coordtrans}
\ee 
and choosing 
\be
M(u)=m(v)=m(-u) \, .  \label{masscond}
\ee 
The apparently opposite arrows of time on either side of the MTT are okay. Keep in mind that these are just the induced coordinates on a spacelike surface. Thus if the associated coordinate vectors point in opposite directions this is not physically significant. 

For comparison with the earlier cases, the expansion parameter may be found kinematically from the tangent vector to the common horizon. 
Respectively on the ingoing and outgoing sides:
\be
\mathcal{V} = \frac{\partial}{\partial v} + \dot{m} \frac{\partial}{\partial r}  =  \frac{\partial}{\partial r} - m' \frac{\partial}{\partial u} \, ,
\ee
so 
\be
C = \dot{m} 
\ee
and 
\be
\epsilon^2 = 2 \dot{m} \, .
\ee
Thus as we already saw in (\ref{hin}) and (\ref{hout}) the common horizon is spacelike. This is the same as for non-extremal evolution
and as before there will be a (finite) discontinuity in the evolution parameter if the derivative of the mass function is also discontinuous.

%The evolution equation (\ref{C2}) is degenerate and so the evolution of the common horizon is found kinematically from 
%the tangent vector to the common  horizon. Respectively on the ingoing and outgoing sides:
%\be
%\mathcal{V} = \frac{\partial}{\partial v} + \dot{m} \frac{\partial}{\partial r}  =  \frac{\partial}{\partial r} - m' \frac{\partial}{\partial u} \, ,
%\ee
%so 
%\be
%C = \dot{m} 
%\ee
%and 
%\be
%\epsilon^2 = 2 \dot{m} \, . 
%\ee
%This common horizon is spacelike when evolving (and so a dynamical horizon) and otherwise null. This is the same as for non-extremal
%evolution. As before there will be a (finite) discontinuity in the evolution parameter if the derivative of the
%mass function is also discontinuous. 

Unfortunately, a thin shell singularity is imposed by this matching. We calculate this by first finding the extrinsic curvatures of the MTT. It
is most convenient to calculate and express them with respect to a tetrad tied to the geometry of the horizon. In the two coordinate systems we have
\bea
\hat{e}_{(0)} &=& \frac{1}{\sqrt{2\dot{m}}} \frac{\partial}{\partial v} \!- \!\sqrt{\frac{\dot{m}}{2}} \frac{\partial}{\partial r\subs{\!in}}  = \frac{1}{\sqrt{2\dot{m}}} \frac{\partial}{\partial u}\! +\! \sqrt{\frac{\dot{m}}{2}} \frac{\partial}{\partial r\subs{\!out}} \\
\hat{e}_{(1)} &=& \frac{1}{\sqrt{2 \dot{m}}} \frac{\partial}{\partial v} \!+\! \sqrt{\frac{\dot{m}}{2}} \frac{\partial}{\partial r\subs{\!in}}= \frac{-1}{\sqrt{2\dot{m}}} \frac{\partial}{\partial u} \!+\! \sqrt{\frac{\dot{m}}{2}} \frac{\partial}{\partial r\subs{\!out}} \\
\hat{e}_{(2)} &=& \frac{1}{m} \frac{\partial}{\partial \theta}   \\
\hat{e}_{(3)} &=&  \frac{1}{m\sin \!\theta} \frac{\partial}{\partial \theta} \, . 
\eea
where $\hat{e}_0$ is the future-oriented unit timelike normal, $\hat{e}_1$ is the outward-pointing unit radial vector (tangent to the MTT) and $\hat{e}_{(2)}$ and $\hat{e}_{(3)}$ are
the unit angular vectors. In the usual way the corresponding one-forms will be written with raised tetrad labels: $\hat{e}^{(0)}$, $\hat{e}^{(1)}$, $\hat{e}^{(2)}$ and $\hat{e}^{(3)}$. 

Then the extrinsic curvatures $K_{(i)(j)}=\hat{e}_{(i)}^a \hat{e}_{(j)}^b \nabla_a \tilde{e}^{(0)}_b$ ($i$ and $j$ only run from 1 to 3) for the infalling (accreting) and outward radiating sides are
\bea
K\sups{in}& = &  - \frac{\ddot{m}}{(2m)^{3/2}} (\hat{e}^{(1)}\otimes \hat{e}^{(1)}) \\
& &  - \frac{1}{m}\sqrt{\frac{\dot{m}}{2}} (\hat{e}^{(2)}\otimes \hat{e}^{(2)} + \hat{e}^{(3)}\otimes \hat{e}^{(3)}) \nonumber
\eea
and
\bea
K\sups{out}  & = &  \frac{\ddot{m}}{(2m)^{3/2}} (\hat{e}^{(1)}\otimes \hat{e}^{(1)}) \\
& & + \frac{1}{m}\sqrt{\frac{\dot{m}}{2}} (\hat{e}^{(2)}\otimes \hat{e}^{(2)} + \hat{e}^{(3)}\otimes \hat{e}^{(3)}) \, . \nonumber
\eea
That is $K\sups{in} = - K\sups{out}$ and so there will be a thin shell singularity along the MTT.

This seems to contradict \cite{Ori1991} where for the non-extremal cases it was stated (but not shown) that the extrinsic curvatures match and so  
there is no thin shell. However in our extreme case, the negative sign is quite intuitive. By (\ref{coordtrans}) and (\ref{masscond}) the two sides of the 
MTT are time-reversed images of each other and so it is not surprising that the extrinsic curvatures, which are essentially time derivatives of the 
intrinsic metric, have opposite signs.  

We can then calculate the form of the stress-tensor for that thin shell:
\be
S_{ij} = - \frac{1}{8 \pi} \left((K\sups{out}_{ij} - K\sups{in}_{ij}) - (K\sups{out} - K\sups{in}) h_{ij} \right) \; . \\
\ee
That is 
\bea
S &=& - \frac{1}{2\pi m} \sqrt{\frac{\dot{m}}{2}}(\hat{e}^{(1)}\otimes \hat{e}^{(1)}) \label{S} \\
& & - \frac{m \ddot{m} + 2 \dot{m}^2}{4 \pi m (2 \dot{m})^{3/2}} (\hat{e}^{(2)}\otimes \hat{e}^{(2)} + \hat{e}^{(3)}\otimes \hat{e}^{(3)}) \, . \nonumber
\eea
Thus there is instantaneous stress at the transition. If $\ddot{m} = 0$ it is an isotropic  (negative) pressure otherwise it is anisotropic. 

Note however that this stress is strangely innocuous. By (\ref{TinLL}), (\ref{TinLN}), (\ref{ToutNN}) and (\ref{ToutLN}) the bulk part of the 
stress-energy tensor is continuous across the MTT: there are no lasting jumps in either energy, momentum or stress density. This may seem
surprising given the appearance of FIG.~\ref{DynEx} (and the fact that the electric current (\ref{EMflow}) certainly does discontinuously 
change direction there). The root of this behaviour can be found in (\ref{TinLL}) and (\ref{ToutNN}) 
which both vanish on the MTT.  The mass-energy density of the dust is exactly balanced by the (negative) electromagnetic potential 
energy on the MTT (see the discussion in footnote \ref{Foot5}) and so the net flow of mass-energy vanishes at the MTT.

These results can be cross-checked from a direct analysis of the junction conditions. 
Following \cite{Barrabes:1991ng} the change in momentum of the matter across the shell is
\be
 \hat{e}_{(k)}^a (T\sups{out}_{ab} - T\sups{in}_{ab}) \hat{e}^b_{(0)}=  \hat{e}^j_{(k)} D_i S^i_{\phantom{i}j} 
\ee
where $D_i$ is the covariant derivative compatible with $h_{ij}$. We have already noted that the left-hand side vanishes but one can also 
directly calculate the right-hand side using (\ref{hin}) and (\ref{S}) and find that it also vanishes (as it should). Similarly the change in the
energy density across the shell is
\be
(T\sups{out}_{ab} - T\sups{in}_{ab}) \hat{e}^a_{(0)} \hat{e}^b_{(0)} = \frac{1}{2} \left(K\sups{in}_{ij} + K\sups{out}_{ij} \right) S^{ij} \,. 
\ee
Above we noted that there is no change in energy density and this can easily be confirmed on the right-hand side since $K\sups{in}_{ij} = - K\sups{out}_{ij}$. 

 Thus while there is a thin-shell singularity imposed by the matching,  it does not appear to have any dramatic physical impact. One might
 speculate that this is what happens when (not-very-physical) charged null dust turns around: it is not the cause of the change in direction but 
 rather the effect. This deserves some further investigation, especially in light of the tension with \cite{Ori1991} which did not find a thin shell 
 in non-extremal cases, however that is beyond the scope
 of this paper and will be left for a future investigation.

Finally, before leaving this spacetime we note the following features. As can be seen in FIG.~\ref{DynEx} the 
inner MTT remains forever null and isolated. No matter ever reaches that surface but instead is bounced back out by the 
combined electric charge of the original black hole and accreted charged dust. 
Further there are no trapped (or anti-trapped) surfaces anywhere in this spacetime.  

\section{Evolutions for scalar fields}
\label{scalar}

We now consider how a massless and uncharged scalar field drives the evolution of initially extremal RN horizons. 
While spacetimes with scalar fields are not exactly solvable, they have been extensively studied both numerically and analytically. 

Most recently much of this interest has focussed on extremal instabilities. While the initial work was in the test field  approximation, 
Murata, Reall and Tanahashi  \cite{Murata:2013daa} (henceforth MRT) have numerically studied the evolution of 
spherically symmetric scalar fields around an initially extremal RN black holes in massless Klein-Gordon Maxwell-Einstein theory.  
We will compare our results to that paper.

\subsection{Evolution from extremal MTTs}
\label{exit2}

For a massless scalar field,  (\ref{Cscalar}) gives:
%
%
%
% the case of a (still spherically symmetric) uncharged and massless scalar field evolving around
%an initially extreme RN horizon that obeys the usual Klein-Gordon equation:
%\be
%\nabla^2 \varphi = 0 \, . 
%\ee
% This is the matter field considered in much of the extremal horizon instability literature. Dynamical exact solutions do not exist for this type of matter however
% we can use the machinery of earlier sections to gain some insight into horizon evolutions in these spacetimes. 
%
%The stress-energy tensor for $\varphi$ plus electromagnetism is
%\be
%T_{ab} =  \frac{1}{4 \pi} \left(\nabla_a \varphi \nabla_b \varphi - \frac{1}{2} (\nabla_c \nabla^c \varphi) g_{ab} \right) + T^{\mbox{\tiny{EM}}}_{ab} \, . 
%\ee
%Thus the expansion parameter  is
\be
\epsilon^2 =\frac{8r^4 (\Lie_\ell \varphi)^2}{r^2 - q^2} \, , 
\ee
where $q$ is constant since there is no means of propagating charge in these spacetimes. 
Evolutions are driven by non-zero $\Lie_\ell \varphi$ and we can apply the intuition gained for null dust to understand
the possibilities here. 

We begin with non-extremal MTTs. 
As usual we have $r \geq q$ for a strictly stably outermost MOTs with $r=q$ for extremality ($\Lie_n \tl = 0$). A subextremal 
MTT will then be null and isolated if $\Lie_\ell \varphi = 0$ and otherwise be dynamical, spacelike and expanding. Correspondingly 
inner horizon MTTs (which are in some sense superextremal) will be null and isolated if $\Lie_\ell \varphi = 0$ and otherwise 
contracting timelike membranes. Thus  any evolution should further separate the geometric horizons and 
drive the black hole farther away from extremality. 

Departures from extremality should be very similar to those which we explored for Vaidya RN. 
The value of the initial expansion will depend on the limiting value of $C$: in principle it could be zero, finite or infinite depending 
on how quickly $\Lie_\ell \varphi$ departs from zero. However whatever the details, the extremal MTT 
splits to become an inner timelike membrane and outer dynamical horizon as in the schematic FIG.~\ref{ExExit}. 

At least for the outer MTT this is in accord with the observations of MRT.  In their paper they study
the evolution of both outgoing and ingoing scalar wave packets and find that generically in both cases the outer horizon settles 
down to a non-extreme black hole. This agrees with our conclusions from (\ref{Cscalar}). If $\Lie_\ell \varphi \neq 0$ at any time then the horizon 
will exit extremality.  For the inner horizon things are a bit more complicated as there are other potential instabilities there. Dealing with these are beyond 
the scope of the current paper and we refer the reader to MRT for a detailed discussion (including evidence that dynamic inner horizons are actually  more stable 
than time-independent ones). 

Returning to the outer MTT, keep in mind that since these matter fields cannot carry charge, $q$ will remain constant throughout any evolution. Thus
dynamical extremal MTTs like those discussed in the previous section are impossible for these spacetimes.

\subsection{Dynamically extremal event horizons}
\label{ExEv2}

However there is an alternate notion of extremality based on causal rather then geometric horizons. A black hole spacetime can be said to be
extremal if the event horizon contains  no trapped surfaces\cite{Israel_ThirdLaw}.

For finely tuned outgoing scalar wave packets, MRT construct examples of such event horizons. 
%
%Up until now we have always assumed that there is an MTT in the spacetime but MRT consider another situation.
% For finely tuned outgoing wave 
%packets they construct examples of spacetimes with an event horizon but no trapped surfaces. 
In their examples, the scalar field accretes onto 
a charged central singularity so that the spacetime asymptotically approaches extremal RN.  They show that are no trapped surfaces anywhere in this 
spacetime and so the event horizon may be considered dynamically extremal. %That event horizon should then clothe at least some of the singularity. 
 
These spacetimes also do not contain any spherical MOTS.
% 
%Not only do these spacetimes not contain spherical trapped surfaces they also do not contain spherical MOTS. 
This follows from: i) for any MOTS $r_{\mbox{\tiny{MOTS}}} \geq q$ \cite{Dain:2011kb}, ii) MOTS are necessarily inside event horizons,  iii) in these 
spacetimes the event horizon only approaches $q$ asymptotically (from below) and (iv) the expansion $\tn$ of the ingoing radial null geodesics is
negative. Thus all surfaces of constant $r$ in the casual future of the event horizon have $r < q$ and so cannot be MOTS.

To get a better intuition for the structure of these dynamically extremal event horizon spacetimes let us return to our much simpler 
null dust spacetimes and consider the analogous situation there. This will also help to motivate a further property of these spacetimes:
they contain naked singularities.  

%
%While the MRT examples are fairly complicated, exact solution analogues of their dynamically extremal event horizons may be constructed 
%in our null dust spacetimes. 
%To get a better intution for this alternate notion of dynamical extremality we first consider the analogous situation for null dust spacetimes. 
Fix $\dot{q} = 0$, $m \leq q$ and $\lim_{v \rightarrow \infty} m = q$. Then there are two  possibilities depending on whether or not the limit is 
achieved. In 
FIG.~\ref{DEfN}a) $m(v)$ asymptotes towards $q$ but doesn't achieve its limit. Thus no MOTS forms but there is still an event horizon: the null surface that
asymptotes to $r=q$ as $v \rightarrow \infty$. FIG.~\ref{DEfN}b) achieves the limit at finite $v$ at which point we switch $\dot{m} = 0$. 
Beyond that point there is an extremal isolated MOTS and this can be evolved backwards to find the full event horizon. 
Note that in both cases there is a singularity which is  visible both inside and outside the horizon. 

\begin{figure}
\includegraphics{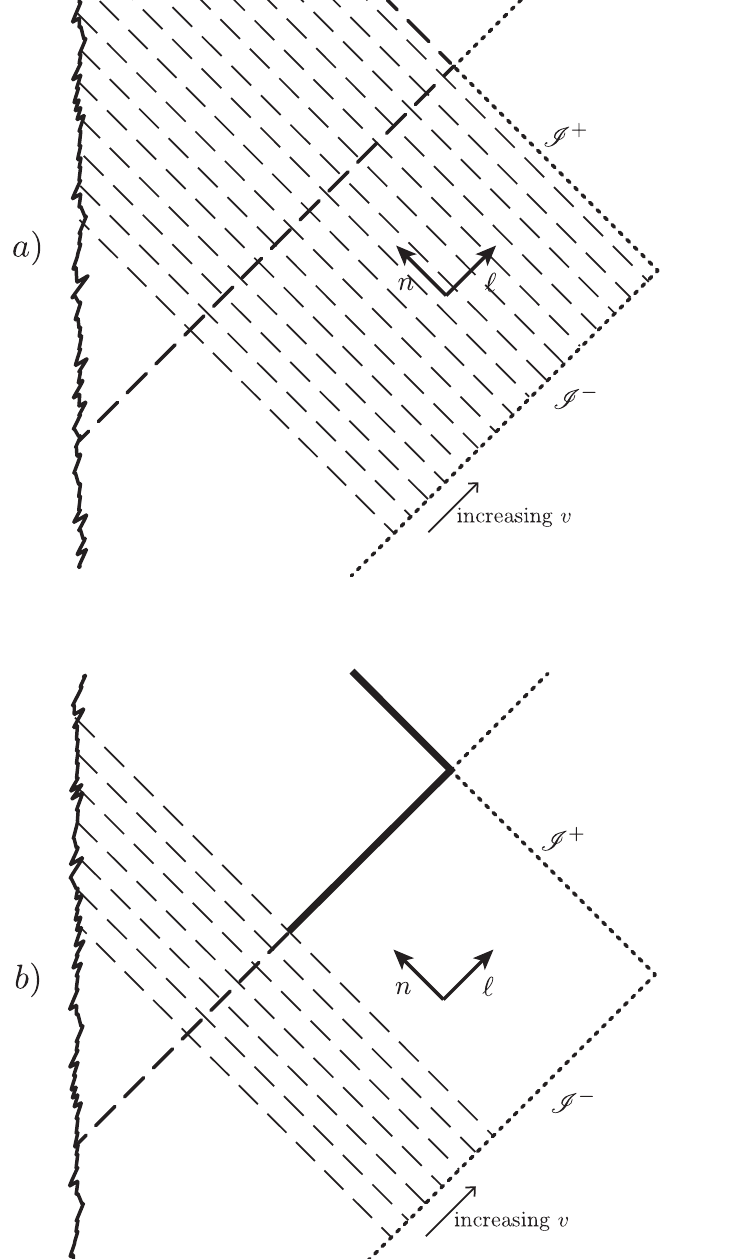}
\caption{Finely tuned evolution of an initially naked singularity into an extremal black hole for null dust spacetimes. 
Long dashed lines are event horizons while heavy solid lines are MTTs. 
Subfigure a) is in the spirit of \cite{Murata:2013daa} with neither trapped nor marginally trapped surfaces. In this case
$m(v) < m_o$ throughout but $\lim_{v\rightarrow \infty} m(v)= m_o$. However in subfigure b) the limit is achieved
and so a MOTS forms. In both cases there is a naked singularity. }
\label{DEfN}
\end{figure}

In fact we argue that such naked singularities are not a special property of the null dust spacetimes but are generic for 
dynamically extremal black hole spacetimes which are charged but have no charge carrying field (and so are also present in MRT). 
In such a spacetime, there is always a singularity at $r=0$ (thanks to the unchanging non-zero electric charge at the origin) and so the only question
is whether the event horizon can fully clothe that singularity. It can't: the Raychaudhuri equation along with the null energy condition implies that all outward moving null shells (this includes any 
event horizon) necessarily emerge from the origin. Null rays outside the event horizon also originate from the singularity and 
so it is visible to observers at infinity. 

It is unclear whether or not naked singularities are a generic feature of dynamically extremal event horizon spacetimes.
If one allows other charge carrying matter fields to be active in the past during the formation of the singularity, then clothing might be possible. 
However even if that is not the case (it seems likely that electrical repulsion would forbid the formation of a $q>m$ singularity even
if charge carrying fields are allowed) for a different choice of matter fields naked singularities might be avoidable.\footnote{A
scenario for the construction of such a spacetime might follow from an example by Williams \cite{Williams10}. For carefully chosen
massive scalar field characteristic initial data representing the interior of a black hole (one of the null surfaces is the event horizon) 
she demonstrated that there could be no (spherically symmetric) MTT asymptotic to any event horizon.
This implies that there are also no (spherically symmetric) trapped surfaces
close to the event horizon. Thus such a spacetime would be at least locally extremal in the sense of  \cite{Israel_ThirdLaw}.
Note however that her construction neither excludes the possibility of trapped surfaces deeper inside the black hole nor a naked singularity in the 
causal past. However, it at least seems possible in principle that 
such an event horizon could fully clothe a central singularity that formed at finite time.}

\section{Discussion}
\label{discuss}

We have seen that geometric VRN horizons evolving from extremality demonstrate several interesting behaviours. 

First, they are tripartite. When matter accretes onto an extremal MTT it bifurcates into a pair of MTTs: an inner and
an outer horizon. While this is physically expected, it is something that does not happen during generic black hole
evolutions. In $(3+1)$ general relativity it has been shown that strictly stably outermost MOTS on 
one slice necessarily evolve uniquely into MOTS on future slices\cite{Andersson:2005gq,Andersson:2008up}. 
However it is clear that the spacetimes depicted in FIG.~\ref{ExExit} can be foliated so that a single horizon does evolve into two. 
There is no contradiction: isolated extremal horizons are not strictly stably outermost. 
%With $\Lie_\ell \tl = \Lie_n \tl = 0$ we also have 
%\be
%\Lie_X \tl = 0
%\ee
%for any vector $X = A \ell + B n$ defined by constant $A$ and $B$. This includes spacelike normals. 

In FIG.~\ref{DEfN} we saw another example of an unexpected behaviour: an extremal marginally trapped tube forming instantaneously.  
Again this isn't usually seen for marginally trapped tubes which may weave backwards and forwards through spacetime but do not
generally appear out of nowhere. During gravitational collapse they may appear out of the central singularity as it forms 
and there are known examples of them forming out of or disappearing into shockwave singularities \cite{Tippett:2014raa}
however as far as we know this kind of appearance ``from thin air'' has not be commented on before. Once more the 
strangely behaving horizon is extremal and so not subject to the usual theorems for strictly stably outermost horizons. 
With this perspective it is not so different from apparent horizon ``jumps'' like the one depicted at $t_B$ in 
FIG.~\ref{MTT_Figure}. There it is also the case that $\Lie_{\hat{r}} \tl = 0$ at the moment of appearance (where $\hat{r}^a$ is the 
radial tangent to the slice).

The transition from isolation can be more dramatic that that for a non-extremal horizon. Linear
($k=1$) accretion will cause a kink in the horizon and jump in the evolution parameter for any MTT (there is a
corresponding discontinuity in the density of accreting matter) however for an initially extremal horizon it generates an 
instantaneous transition to a null membrane (with $\epsilon^2 = \pm \infty$). Even with a continuous matter 
distribution there can be kinks in the MTT (for $k=2$ matter). The extremality introduces an extra level of possible
discontinuity due to the form of the expansion parameter (\ref{Cpm}). Similar results follow for transitions driven by 
scalar fields. 

We also examined two notions of dynamical extreme spacetimes. The first was for a dynamic extremal MTT. There infalling $q=m$ matter
generates a pair of extremal MTTs (one isolated and the other dynamic). Constructing these examples in exact form required the careful 
use of spacetime surgery to connnect ingoing and outgoing Vaidya solutions. While the bulk part of the stress-energy tensor is continuous across
the junction, the extrinsic curvatures on the two sides of the junction do not match (they differ by a negative sign) and so an instantaneous 
thin shell of matter makes an appearance along the junction. Its physical significance is not entirely clear and deserves further investigaton. 
While these
evolutions are of doubtful physical importance being both finely tuned and driven by unrealistic matter, they remain as interesting 
examples of possible behaviours that are not forbidden by the energy conditions. 

The second notion of dynamical extremality, proposed in \cite{Murata:2013daa}, is a spacetime with an event horizon but no trapped or even marginally trapped surfaces.
We considered such situations for null dust spacetimes and saw that they contain naked singularities. From more general considerations it also appears that any other such spacetime 
with a fixed central electric charge will be nakedly singular. In such spacetimes dynamical extreme event horizons emerge from naked singularities. However we also noted that
for other matter fields this may not be the case. 
%For example  a massive scalar spacetime with an event horizon but no MTT was constructed in \cite{Williams10}. 

\section*{Acknowledgements} Thanks to Hari Kunduri and Andrey Shoom for useful discussions about this work and Harvey Reall for his comments on an earlier version of this paper.  IB was supported by NSERC Discovery Grant 261429-2013.  

\appendix

\bibliography{FromExtremality_PRDEdit}

\end{document}